\documentclass[sigconf]{acmart}
\AtBeginDocument{%
  }

\usepackage{csquotes}
\usepackage{tabularx}
\usepackage{multirow}
\usepackage[table]{xcolor}

\setcopyright{acmlicensed}
\copyrightyear{2025}
\acmYear{2025}
\acmDOI{}

\acmConference[SciSoc LLM Workshop]{KDD'25 SciSoc LLM Workshop Large Language Models for Scientific and Societal Advances}{2025}{Toronto, CA}
\acmISBN{}

\begin{document}

\title{Dissecting Judicial Reasoning in U.S. Copyright Damage Awards: A Discourse-Based LLM Analysis}

\author{Pei-Chi Lo}
\email{pclo@mis.nsysu.edu.tw}
\affiliation{%
  \department{Department of Information Management}
  \institution{National Sun Yat-sen University}
  \city{Kaohsiung City}
  \country{Taiwan}
}

\author{Thomas Y. Lu}
\email{lewislu@mail.nsysu.edu.tw}
\affiliation{%
  \department{Department of Business Management}
  \institution{National Sun Yat-sen University}
  \city{Kaohsiung City}
  \country{Taiwan}
}
\renewcommand{\shortauthors}{}

\begin{abstract}
    Judicial reasoning in copyright damage awards poses a core challenge for computational legal analysis: although federal courts follow the 1976 Copyright Act, their interpretations and factor weightings vary widely across jurisdictions. This inconsistency creates unpredictability for litigants and obscures the empirical basis of legal decisions. This research introduces a novel discourse-based Large Language Model (LLM) methodology that integrates Rhetorical Structure Theory (RST) with an agentic workflow to extract and quantify previously opaque reasoning patterns from judicial opinions.
    Our framework addresses a major gap in empirical legal scholarship by parsing opinions into hierarchical discourse structures and using a three-stage pipeline—Dataset Construction, Discourse Analysism and Agentic Feature Extraction. This pipeline identifies reasoning components and extract feature labels with corresponding discourse subtrees. In analyzing copyright damage rulings, we show that discourse-augmented LLM analysis outperforms traditional methods while uncovering unquantified variations in factor weighting across circuits.
    These findings offer both methodological advances in computational legal analysis and practical insights into judicial reasoning, with implications for legal practitioners seeking predictive tools, scholars studying legal principle application, and policymakers confronting inconsistencies in copyright law.
\end{abstract}


\keywords{Judicial Reasoning, Copyright Damage, Information Retrieval, Large Language Models}

\maketitle

\section{Introduction}
    \subsection{Motivation}
    The systematic analysis of judicial reasoning represents one of the most complex challenges in computational legal studies. While statutory frameworks provide seemingly uniform guidance for judicial decision-making, the translation of legal principles into specific case outcomes involves intricate reasoning processes that vary substantially across jurisdictions, time periods, and individual judicial perspectives. This variation, while reflecting the adaptive nature of legal interpretation, creates significant challenges for litigants seeking to predict case outcomes, scholars attempting to understand legal system dynamics, and policymakers evaluating the effectiveness of statutory frameworks.
    
    Copyright damage awards under the U.S. Copyright Act of 1976 exemplify this analytical challenge with particular clarity. The Act establishes two primary calculation methodologies: actual damages and profits, which compensate copyright owners for demonstrable losses and disgorge infringers' unlawful gains; and statutory damages, which permit plaintiffs to elect predetermined amounts ranging from \$750 to \$30,000 per infringed work, with potential enhancement to \$150,000 for willful violations. While this dual framework appears to provide clear guidance, judicial application reveals substantial discretionary variation that profoundly impacts litigation outcomes.

    \subsection{Empirical Evidence of Judicial Variation}
    Systematic examination of federal circuit court decisions reveals striking inconsistencies in judicial reasoning approaches. In \textit{Christopher Phelps \& Associates, LLC v. Galloway}~\cite{phelps2007}, Judge Niemeyer emphasized market value assessment and reasonable royalty calculations as foundational elements for actual damages, stating that such damages ``are measured according to market value, which means what a willing buyer would have been reasonably required to pay a willing seller for the copyright holder's work.''
    Conversely, in \textit{Lucky Break Wishbone Corp. v. Sears, Roebuck Co.}~\cite{lucky2010},  Judge Zilly focused extensively on burden-shifting frameworks and causation requirements, providing minimal attention to market valuation concepts despite addressing similar infringement scenarios. 

    This variation extends prominently to statutory damage determinations. In \textit{Jett v. Ficara}~\cite{jett2007}, Judge Pitman articulated a comprehensive framework incorporating ``the expenses saved and profits reaped by the defendants, the revenues lost by the plaintiffs, the value of the copyright, the deterrent effect of the award on other potential infringers, and factors relating to individual culpability.'' However, in \textit{ Microsoft Corp. v. Gonzales Civil.}~\cite{micro2007}, Judge Kugler emphasized defendant intent and behavior as ``the foremost consideration,'' largely subordinating economic factors to behavioral analysis. 

    These documented variations highlight fundamental limitations in traditional approaches to legal analysis. Conventional doctrinal scholarship, while providing valuable theoretical frameworks, struggles to quantify reasoning pattern variations at the scale necessary for systematic understanding. Manual case analysis, though detailed, remains resource-intensive and potentially subject to researcher bias. Most critically, existing methodologies lack the computational capacity to extract and analyze reasoning patterns across large corpora of judicial opinions, limiting our empirical understanding of how legal principles translate into practical outcomes.

    \subsection{Research Questions}
    This research addresses these limitations through a novel computational methodology that combines discourse analysis with advanced Large Language Model capabilities. Our approach leverages Rhetorical Structure Theory (RST)~\cite{rst} to decompose judicial opinions into hierarchical representations of argumentative structure, then employs a sophisticated three-stage LLM pipeline to extract, categorize, and analyze reasoning patterns with unprecedented precision and scale. 
    Our research addresses two fundamental research questions:

     \textbf{RQ1: Methodological Validation.} Can discourse-based LLM analysis accurately extract complex reasoning patterns from judicial opinions at scale, demonstrating superior performance compared to traditional computational approaches?
     
     \textbf{RQ2: Empirical Discovery.} How can discourse-augmented LLM analysis reveal previously undetectable patterns in judicial reasoning, particularly regarding factor prioritization and jurisdictional variation in copyright damage determinations?
    
    To the best of our knowledge, this study represents the first systematic attempt to analyze the judicial analytical frameworks used in determining damage awards in copyright infringement cases. By addressing these questions through rigorous computational analysis of federal court decisions, this research establishes foundations for enhanced predictive legal analytics, improved understanding of judicial decision-making processes, and evidence-based evaluation of statutory framework effectiveness. The implications extend beyond copyright law to encompass broader questions about judicial consistency, legal system transparency, and the role of empirical evidence in legal scholarship and policy development.
    

    \begin{figure*}[t!]
        \centering
        \includegraphics[width=\linewidth]{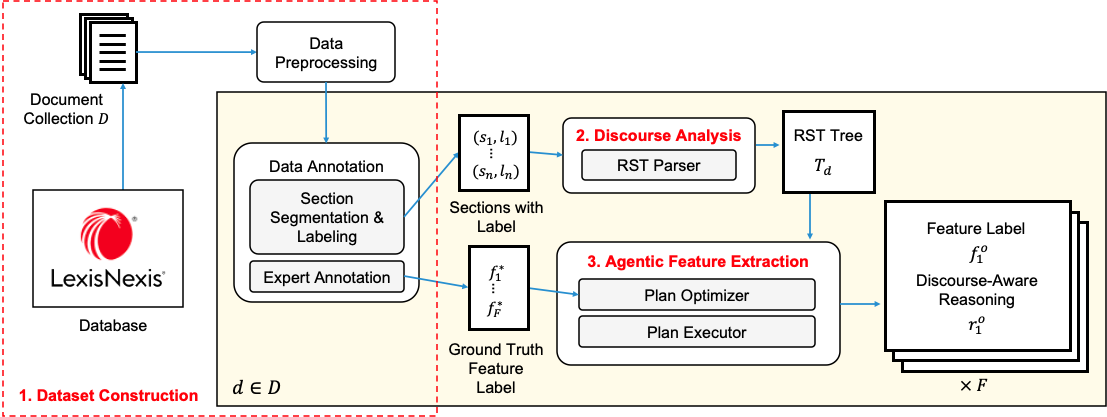}
        \caption{Research Framework }
        \label{fig:FW}
    \end{figure*}
\section{Related Works}
    \subsection{Copyright Damage Analysis}

    The literature on copyright damages has evolved distinctly across different time periods. Works before 2000 were predominantly doctrinal, focusing on introducing the details of the 1976 Copyright Act and debating core concepts of copyright damages calculations~\cite{coleman1993copyright}. Scholarship from 2000-2010 shifted emphasis toward addressing statutory damage calculation in the context of Internet prevalence. After 2010, discussions continued to critique the statutory damage regime, with scholars also beginning to apply concepts from other legal fields (such as patent law) to copyright damage analysis~\cite{barker2004grossly}.

    A significant limitation throughout this evolution has been the absence of empirical research until Professor Depoorter's pioneering work, which employed descriptive analysis to examine copyright infringement cases between 2005 and 2008~\cite{depoorter2019copyright}. However, even this research did not empirically investigate what factors might affect the ultimate award amounts or how those factors influence results under different calculation methods.

    This represents the central concern for plaintiffs and defendants in copyright litigation and forms the primary research question addressed in this paper. By analyzing judicial reasoning through large language models, this research bridges the gap between doctrinal scholarship and empirical analysis, offering insight into how different factors influence copyright damage awards across jurisdictions.

    \subsection{LLMs for Legal Analysis}
    Recent advances in applying Large Language Models (LLMs) to legal text have demonstrated promising capabilities in several key areas. Legal Judgment Prediction (LJP) tasks have shown that LLMs can analyze case facts to predict outcomes, charges, and applicable laws~\cite{tong_legal_2024,xu_distinguish_2024,yue_circumstance-aware_2024}. Deng et al. formalized Syllogistic Legal Judgment Analysis (SLJA) to break down legal reasoning into component tasks~\cite{deng_syllogistic_2023}, while Li et al. proposed chain prompt reasoning to uncover logical structures in legal texts~\cite{li_basis_2025}. These approaches inform our methodology for extracting reasoning patterns from copyright opinions.
    
    Several researchers have enhanced LLM performance through knowledge integration. Shu et al. developed LawLLM specifically for the US legal domain, performing tasks like Similar Case Retrieval (SCR) and precedent recommendation~\cite{shu_lawllm_2024}. Zhang et al. demonstrated improved case retrieval by incorporating event-based knowledge through heterogeneous knowledge graphs, which inspires our factor identification approach~\cite{zhang_event_2024}.
    
    Despite these advances, significant limitations exist in current LLM applications to legal analysis. Most empirical research uses Chinese legal datasets~\cite{deng_syllogistic_2023,li_basis_2025}, with limited work on US doctrine. Interpretability remains a challenge, with many models functioning as ``black boxes''~\cite{lai_large_2024}. Additionally, most work focuses on predicting discrete outcomes rather than analyzing nuanced reasoning processes that lead to those outcomes. Our research addresses this gap by focusing on LLM-assisted fine-grained legal judgment reasoning analysis, specifically in copyright infringement cases.

\section{Methodology}
    
        Our research employs a three-stage framework to analyze judicial reasoning in copyright damage awards, as illustrated in Figure~\ref{fig:FW}.
        \begin{enumerate}
            \item \textbf{Dataset Construction:} We begin by collecting judicial opinions from the LexisNexis Database ($D$) containing copyright damage discussions. Each document ($d \in D$) undergoes data preprocessing to extract structured content, followed by a dual-layer annotation process. This includes automatic section segmentation and labeling performed by an LLM to identify functional components of judicial opinions, complemented by expert annotation that provides ground truth labels ($f_1^* ... f_p^*$) for damage awards.
            \item \textbf{Discourse Analysis:} For each section with appropriate labels $(s_1, l_1) ... (s_n, l_n)$, we apply Rhetorical Structure Theory (RST) parsing to decompose the text into a hierarchical representation of discourse relationships. This produces RST trees ($T_d$) that capture the logical structure of judicial reasoning, revealing how judges connect various factors in their damage calculations.

            \item \textbf{Agentic Feature Extraction:} We leverage the RST trees and ground truth labels to extract judicial reasoning patterns using a two-component LLM pipeline: (1) a Plan Optimizer that formulates effective prompt to generate feature extraction plans and structures the extraction process, (2) a Plan Executor that execute the plan. For a document $i$, the output is the predicted feature labels ($f^o_1, ..., f^o_F$) with their associated discourse-aware reasoning  ($r_1^o, ..., r_F^o$).
        \end{enumerate}

    \subsection{Dataset Construction}
        Our dataset construction process consists of three sequential phases: (a) data querying, (b) data preprocessing, and (c) data annotation.
    
        \textbf{Data Querying.} We retrieve judicial opinions from the LexisNexis Database using the specific query "Copyright /p Damages award" with the following parameters: statutory provisions 17 U.S.C. § 504, § 505, § 502, and § 501; restricted to Practice Areas \& Topics: Computer \& Internet Law and Copyright Law. This query yields 2,183 cases. Importantly, this initial dataset contains many cases that do not specifically involve copyright infringement with damage awards. Our subsequent filtering processes, detailed later, identify the subset of relevant cases for analysis.
    
    \textbf{Data Preprocessing.} The downloaded LexisNexis Database files are in PDF format, with each file containing approximately 100 cases. We first extract the textual content from these PDFs and separate individual cases. For each case, we systematically extract key metadata including \textit{Case Title}, \textit{Case Number}, \textit{Case Argued Date}, \textit{Case Decided Date}, \textit{Court Name}, \textit{Court District}, \textit{Court Circuit}, and the complete \textit{Judge Opinion}. Additionally, we extract all citations to related cases and statue mentioned within the opinions. This extraction creates a heterogeneous graph structure where nodes represent either legal cases or statute, and edges represent citation relationships between them.
    
    \textbf{Data Annotation.} Our research employs a dual-layered annotation approach, combining expert legal knowledge with computational efficiency to create a reliably labeled dataset that accurately represents judicial reasoning patterns in copyright damage awards:
    \begin{enumerate}
        \item \textbf{Expert Annotation:} A copyright law expert manually reviews and annotates 203 cases, documenting the actual damage awards granted by judges in each case. These expert annotations provide ground truth data for training a classifier that can identify cases directly related to copyright infringement, which forms a critical component of our analysis pipeline. The annotated features include factors that have been shown to relate to the amount of damage awards, such as the defendant's intent, whether fair use was claimed, the type of work infringed, whether the case involved trademark litigation, etc. In this work, we focus on the extraction of ``punitive damage or not.''
        
        \item \textbf{Section Segmentation and Labeling:} Since judicial opinions are typically lengthy and structurally complex, we segment them into functional sections for more effective analysis. We utilize ChatGPT-4o to divide each opinion into clearly defined sections with standardized labels: \textit{Introduction}, \textit{Background Facts}, \textit{Procedural History}, \textit{Analysis of The Infringement}, \textit{Analysis of The Liability}, \textit{Analysis of The Relief and Damages}, and \textit{Order}. These section categories were developed in consultation with the same copyright expert who performed the manual annotations. \\
        While the LLM can identify sections that fall outside these predefined categories by creating labels with the prefix \textit{NEW\_}, we discourage this to maintain consistent structure across the dataset. To verify annotation quality, our copyright expert evaluated ten randomly sampled cases with both LLM and expert annotations, finding that the LLM achieved 92\% accuracy compared to expert-generated section labels. We show the prompt and the definition of each label in the Appendix.
    \end{enumerate}

    \subsection{Discourse Analysis Using Rhetorical Structure Theory (RST)}
    \begin{figure*}[t!]
        \centering
        \includegraphics[width=\linewidth]{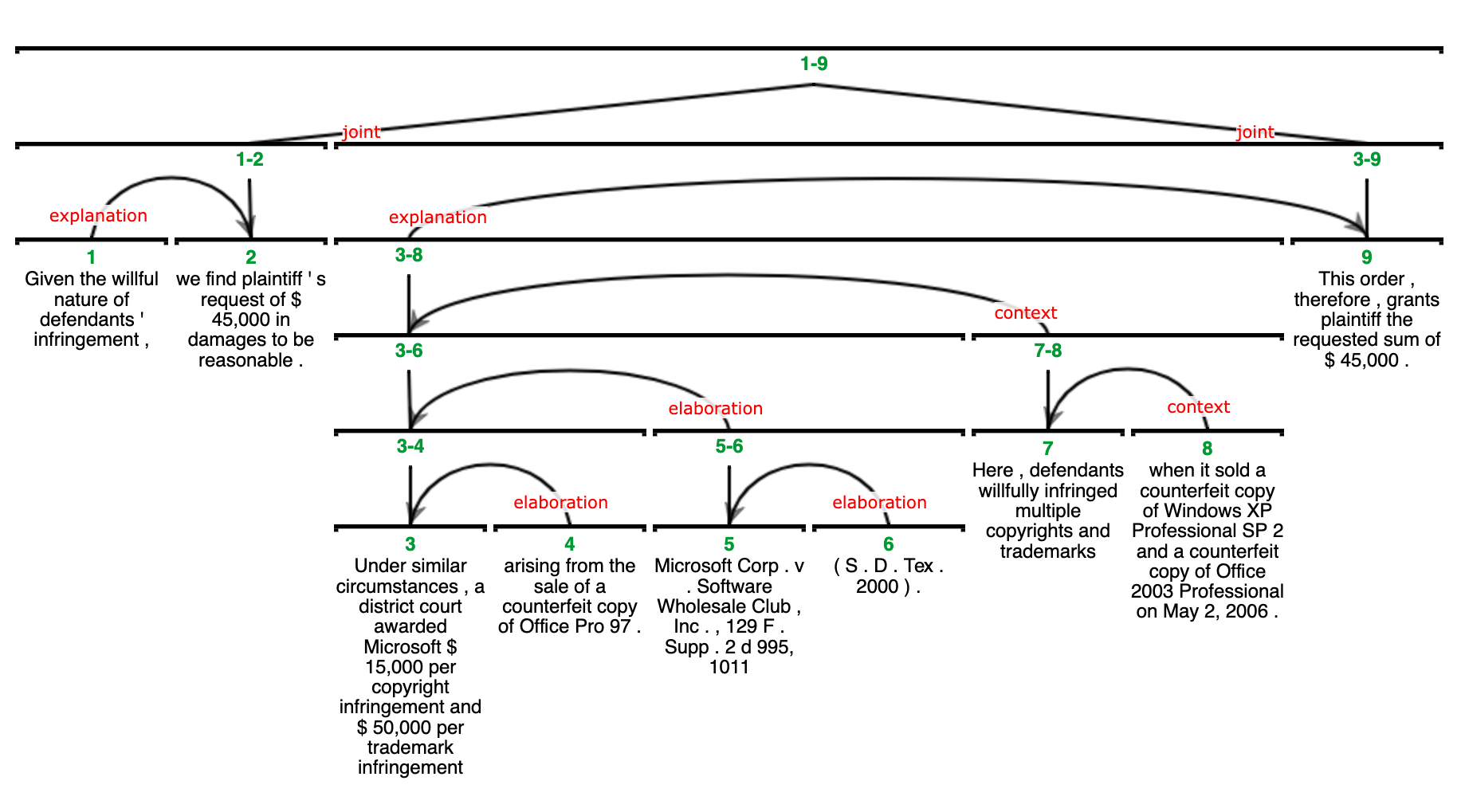}
        \caption{Example RST Tree from \textit{Microsoft Corp. v. E\&M Internet Bookstore, Inc. }~\protect\cite{micro20072}}
        \label{fig:RST_example}
    \end{figure*}


        Rhetorical Structure Theory (RST)~\cite{rst} provides a framework for analyzing text organization by identifying relationships between different text spans, revealing how structure contributes to meaning and coherence. 
        RST defines 23 relations, each describing the constraints on the participating text spans (i.e., nucleus and satellite) and the intended effect on the reader. For instance, in the relation \textit{elaboration}, the satellite presents additional details about the situation or subject matter presented in or inferentially accessible from the nucleus. The intended effect is for the reader to recognize these additional details and identify the specific element being elaborated upon. In this work, we leverage RST to analyze the discourse structure of judicial opinions.
        
        Figure~\ref{fig:RST_example} illustrates an RST tree from the case \textit{Microsoft Corp. v. E\&M Internet Bookstore, Inc.} This tree comprises 9 text spans that demonstrate the judge's reasoning for awarding the plaintiff \$45,000 in damages. The tree shows two primary justifications for this award: (1) the willful nature of the defendants' infringement (spans 1-2), and (2) precedent from similar cases (spans 3-8).
        
        In the first branch, span 1 serves as an explanation for span 2, establishing that the willful nature of infringement makes the \$45,000 request reasonable. The second branch provides legal precedent through citation of \textit{Microsoft Corp. v. Software Wholesale Club, Inc., 129 F. Supp. 2d 995, 1011 (S.D. Tex. 2000)}, where similar circumstances led to substantial damage awards. Spans 7-8 offer specific context about the defendants' actions, noting they ``willfully infringed multiple copyrights and trademarks when it sold a counterfeit copy of Windows XP Professional SP 2 and a counterfeit copy of Office 2003 Professional on May 2, 2006.''
       
        This structured representation allows us to systematically identify the analytical framework judges use when determining copyright damage awards. We parse all cases in our dataset using the RST parser developed by \cite{rst_parser}.

        To enable LLMs to interpret the rhetorical meaning behind a specific text span using RST information, we propose a prompting method that verbalizes an RST tree into natural language. 
        Given a target text span, we first construct a linearized path that traverses from this span back to the root of the RST tree. We then ask the LLM to convert this path into a natural language paragraph, explaining how the target span contributes to the larger structure of the legal opinion. By performing this linearization and verbalization of RST paths, LLMs can more effectively capture the discourse structure of judicial reasoning.

    \subsection{Agentic Feature Extraction}
        To systematically identify and explain the judge's analytical framework in determining copyright damage awards, we implement an agentic LLM-based workflow that automatically extracts feature labels representing components of judicial reasoning and their associated reasoning processes. This two-stage pipeline leverages LLMs to identify reasoning patterns that would be prohibitively resource-intensive to extract manually.
        
        \begin{enumerate}
            \item \textbf{Plan Optimizer.} 
            For each target feature $i$ in our analytical framework, we first generate an initial prompt that asks the LLM to produce a plan for feature extraction. We then provide the LLM with four key inputs: (1) the RST tree $T_d$, representing the discourse structure of the judicial opinion; (2) expert-annotated ground truth examples $f_i^*$ that exemplify the feature in question; (3) plan optimization instructions; and (4) the initial prompt. We implement the automatic prompt optimization method proposed in~\cite{pryzant-etal-2023-automatic}, which first executes the plan and then iteratively refines the plan generation prompts $p_i$ by minimizing the gap between the feature extraction output and the ground truth annotations. If the evaluation determines the extraction is incorrect, $p_i$ is revised to resolve the identified mistake. This iterative refinement continues until a valid extraction is achieved, ensuring high-quality analysis of judicial reasoning patterns across a heterogeneous corpus of judicial opinions.
            
            \item \textbf{Plan Executor.} Using the optimized prompt $p_i$, we query an LLM to generate a feature extraction plan for feature $i$. This plan details how to traverse the RST tree to identify text spans that represent the feature and their relationships to other reasoning elements. We then execute the plan, and output both the predicted feature label $f_i$ and its associated discourse-aware reasoning $r_i^o$, where $r_i^o$ is a subtree extracted from the original RST tree $T_d$ that contains the specific discourse units and rhetorical relations supporting the feature. This subtree captures the logical structure of how judges articulate and justify the particular reasoning element within their broader analysis. This component provides a structured representation of where and how the feature appears in the judicial reasoning, mapping the abstract feature to concrete textual evidence in the opinion.
        \end{enumerate}

        This approach enables us to extract a comprehensive set of features representing different aspects of judicial reasoning in copyright damage cases, including considerations of willfulness, market value assessment, revenue attribution methods, and deterrence objectives. By applying this framework across our corpus, we can quantify how different judges prioritize various factors in their damage calculations and identify patterns that vary by jurisdiction, case type, or time period.
    
\section{Preliminary Results}
    \subsection{Task Formulation}
    To address our two research questions, this study focuses on the feature \textbf{``Whether Punitive Damage Is Considered When Granting Damage Award.''} A punitive component includes damages that go beyond compensation and are intended to punish the defendant or deter future misconduct. This may include: (1) Enhanced statutory damages explicitly awarded under \textit{17 U.S.C. § 504(c)(2)} for willful infringement with a stated purpose of punishment or deterrence; (2) Damages calculated to meaningfully exceed actual harm, when justified by the court as necessary to deter future violations. This feature has been shown as an important factor in deciding the amount of damage award in previous works.
    
    The LLM-based model should generate the following two outputs: (a) predicted label for the input case \{true | false\} indicating whether the judge has considered a punitive component in granting damage award, and (b) the reasoning process. For all experiments, we use OpenAI \textit{ChatGPT-4o-mini} model with temperature set to 0.

    \subsection{RQ1: Methodological Validation}
        To address the first research question, we conduct quantitative analysis to examine the label prediction performance of our proposed discourse-based LLM analysis framework.
        
        \textbf{Dataset.} Due to limited time and effort, we randomly select 50 annotated cases as the dataset in this preliminary experiment. The annotation labels are carefully examined by an expert. We use 15 of the cases for prompt optimization and initial planning, and the remaining 35 cases for testing purposes. The label distribution in the dataset is positive:negative = 4:6.
    
        \textbf{Baselines.}  In addition to our proposed method, we include four baselines in the preliminary experiment:
            \begin{enumerate}
                \item \textsf{Random Guess:} Randomly assigns labels based on the overall label distribution (i.e., a 0.4 probability of predicting positive and 0.6 for negative).
                \item \textsf{Vanilla LLM:} Provides the entire document to the LLM without using any agentic workflow.
                \item \textsf{CoT:} Provides the entire document to the LLM without using any agentic workflow, with zero-shot chain-of-thought prompting.
                \item \textsf{Agentic LLM:} Uses an agentic LLM where the task is divided into steps with reflection and re-planning.
                \item \textsf{Agentic LLM + ToD (Proposed Method):} Uses an agentic LLM with access to the RST discourse tree (i.e., Tree of Discourse, ToD) as part of the workflow.
            \end{enumerate}
        \textbf{Evaluation Metrics.} We report Accuracy (ACC), Precision (PRE), Recall (REC), and F1-score (F1) of the models' performance on the test set.
        
        \textbf{Experiment Results.} We show the performance comparison in Table~\ref{tbl:result}. The results demonstrate that the proposed \textsf{Agentic LLM + ToD} consistently outperforms all baselines across all evaluation metrics. It achieves the highest accuracy (0.771) and F1 score (0.765), indicating a superior balance between precision and recall. Importantly, it also maintains perfect recall (1.000), successfully identifying all instances labeled as containing a punitive component.

        The improvement in precision observed in \textsf{Agentic LLM + ToD} suggests that access to discourse structure information enables the LLM to more effectively distinguish between genuinely punitive reasoning and surface-level cues. While all models exhibit strong recall, this model produces fewer false positives, indicating its ability to recognize subtle rhetorical patterns that are indicative of punitive intent.
        
        The \textsf{CoT} model shows a modest improvement in precision (0.571) over the Vanilla model (0.565), while maintaining identical accuracy (0.714). This suggests that zero-shot CoT prompting introduces a more structured reasoning process, reducing overgeneralization. However, this model still exhibits a tendency to overpredict punitive intent, similar to \textsf{Vanilla LLM} and \textsf{Agnetic LLM}, likely due to the absence of contextual discourse signals.
        
        Both the \textsf{Vanilla LLM} and \textsf{Agnetic LLM} models achieve high recall (1.000 and 0.923, respectively), but at the cost of reduced precision. This indicates a consistent pattern of overpredicting punitive classifications, often triggered by the presence of terms such as ``willful'' or by awards that appear numerically large. These models appear to rely heavily on surface features, such as mentions of \textit{17 U.S.C. § 504(c)(2)}, without fully attending to the legal or rhetorical context in which those features are presented.
        
        The performance disparity between \textsf{CoT} and \textsf{Agnetic LLM} suggests that both agentic prompting (step-by-step decomposition) and intermediate reasoning (chain-of-thought) contribute positively to task performance. However, the superior results achieved by \textsf{Agentic LLM + ToD} demonstrate that incorporating discourse-level rhetorical structure provides a significant additional benefit. The findings underscore the importance of both structured workflow and rhetorical context for improving legal judgment in complex tasks such as identifying punitive components in judicial opinions.

    \begin{table}[t!]
        \centering
        \caption{Preliminary Result}
        \label{tbl:result}
        \begin{tabular}{lcccc}
        \hline
        \textbf{Method} & \textbf{ACC} & \textbf{PRE} & \textbf{REC} & \textbf{F1} \\
        \hline
        Random Baseline & 0.517 & 0.370 & 0.423 & 0.392 \\ 
        Vanilla LLM         & 0.714 & 0.565 & 1.000 & 0.722 \\
        CoT         & 0.714 & 0.571 & 0.923 & 0.706 \\
        Agentic LLM            & 0.657 & 0.522 & 0.923 & 0.667 \\
        Agentic LLM + ToD (Ours)       & \textbf{0.771} & \textbf{0.619} & \textbf{1.000} & \textbf{0.765} \\
        \hline
        \end{tabular}
    \end{table}
    \begin{table*}[hbt!]
        \centering
        \small
        \renewcommand{\arraystretch}{1.2}
        \caption{Case Study Comparison of Predictions and Reasoning Across Methods}
        \label{tbl:case-study}
        \begin{tabularx}{\textwidth}{|c|X|}
        \hline
        \rowcolor{gray!20}
        \textbf{Index} & \textbf{Details and Model Outputs} \\
        \hline
        
        \multirow{6}{*}{1} 
        & \textbf{Case Name:} \textit{IO Group, Inc. v. Jordon~\cite{io2010}} \hfill \textbf{Ground Truth:} T \\
        & \textbf{Excerpt:} Based on these relevant circumstances, an award on the lower end of the scale would have the desired deterrent effect for someone in Defendant's position.
        Accordingly, the Court thinks it just to award statutory damages in the amount of \$ 1,500 per moving picture copyright infringed and \$ 750 per photo trademark infringed, for a total of \$ 21,750. \\ \cline{2-2}
        & \textbf{Vanilla LLM (T):} The court awarded statutory damages under \textit{17 U.S.C. § 504(c)(1)} and considered enhanced statutory damages under \textit{§ 504(c)(2)} for willful infringement, which indicates a punitive component intended to deter future misconduct. The court's reasoning connected the defendant's actions to willfulness and the need for deterrence, thus justifying the punitive nature of the damages. \\
        & \textbf{CoT (F):} The damages awarded were not intended to punish the defendant or deter future misconduct beyond reasonable compensation, as the court opted against enhanced statutory damages despite acknowledging willfulness. \\
        & \textbf{Agentic LLM (F):} The court awarded statutory damages of \$21,750, which is within the standard statutory range and does not include an enhanced amount for willfulness. Although the court acknowledged the possibility of enhanced damages for willful infringement, it ultimately decided against them, stating that the circumstances did not warrant such an increase. The damages awarded do not explicitly or functionally include a punitive component as they are not intended to punish or deter beyond compensating for the actual harm. \\
        & \textbf{Agentic LLM + ToD (T):} The court awarded statutory damages under \textit{17 U.S.C. § 504(c)(1)} and considered enhanced statutory damages under \textit{§ 504(c)(2)} due to the defendant's alleged willful infringement. The court's reasoning explicitly connects the willfulness of the infringement to the potential for punitive damages, indicating that the damages serve both compensatory and punitive purposes. Although the court ultimately decided against enhanced damages, the acknowledgment of willfulness and the potential for increased penalties demonstrate a punitive component in the damages awarded. \\
        \hline
        
        \multirow{6}{*}{2} 
        & \textbf{Case Name:} \textit{Prod. v. Rivan~\cite{rivan}} \hfill \textbf{Ground Truth:} F \\
        & \textbf{Excerpt:} Here, the Court finds that the evidence submitted in support of the motion for default judgment adequately establishes that Defendant's infringement was willful... Thus, at a price of less than \$100.00 per fitness kit, Defendant has generated at most \$4,500 in sales. In light of these facts, the Court will award Plaintiff statutory damages in the amount of \$30,000. \\\cline{2-2}
        & \textbf{Vanilla LLM (T):}  The court awarded \$30,000 in statutory damages for willful copyright infringement, which is explicitly allowed under \textit{17 U.S.C. § 504(c)(2)} for willful infringement. The court's reasoning indicates that the infringement was willful, and the damages awarded exceed the standard statutory range, suggesting a punitive intent to deter future misconduct.\\
        & \textbf{CoT (T):}  The court awarded \$30,000 in statutory damages for willful copyright infringement, which exceeds the actual harm and is intended to punish the defendant and deter future misconduct.\\
        & \textbf{Agentic LLM (T):}  The court awarded \$30,000 in statutory damages for willful copyright infringement, which is within the range allowed under \textit{17 U.S.C. § 504(c)(1)}. However, the court's acknowledgment of the defendant's willful infringement and the discretion to award damages in excess of the standard range indicates a punitive intent to deter future violations, especially given the context of the defendant's continued infringement despite receiving cease and desist letters.\\
        & \textbf{Agentic LLM + ToD (F):}  The court awarded \$30,000 in statutory damages, which is within the standard statutory range for copyright infringement and does not exceed the maximum allowed for willful infringement. Although the defendant's actions were deemed willful, the court did not explicitly connect the damages to a punitive purpose or intent to deter future misconduct, nor did it award damages exceeding the statutory limits that would indicate a punitive component.\\
        \hline
        
        \multirow{6}{*}{3} 
        & \textbf{Case Name:} \textit{Photo Res. Hawai'i, Inc. v. Am. Hawai'i Travel, Inc.~\cite{photo}} \hfill \textbf{Ground Truth:} F \\
        & \textbf{Excerpt:} The Copyright Act provides for statutory damages as follows: ... (2) In a case where... infringement was committed willfully, the court in its discretion may increase the award of statutory damages to a sum of not more than \$150,000. \textit{17 U.S.C. § 504(c)}... In the present case, Plaintiff requests \$18,000 in statutory damages under the Copyright Act. This is three times the amount that Plaintiff would have received had AHTI obtained a license to use the eight photographic works... Plaintiff has established that AHTI wilfully infringed upon Plaintiff's copyrights. This Court therefore finds that Plaintiff's request for \$18,000.00 in statutory damages under the Copyright Act is appropriate under the circumstances. \\\cline{2-2}
        & \textbf{Vanilla LLM (T):}  The court awarded \$18,000 in statutory damages under the Copyright Act, which is three times the amount that would have been received had AHTI obtained a license. This amount exceeds the standard statutory range and is justified by the court as appropriate due to AHTI's willful infringement, indicating a punitive intent to deter future violations.\\
        & \textbf{CoT (T):}  The damages awarded include a punitive component as they are based on the defendant's willful infringement, with the court explicitly connecting the need for higher damages to the intent to punish and deter future misconduct\\
        & \textbf{Agentic LLM (T):}  The court recommended awarding \$18,000 in statutory damages under the Copyright Act, which is three times the licensing fee that would have been charged for the use of the images, indicating a punitive intent to deter future violations. Additionally, the request for \$30,000 in statutory damages under the Digital Millennium Copyright Act was justified by the plaintiff's argument that the defendant's conduct was particularly damaging and warranted a deterrent effect, further supporting the presence of a punitive component.\\
        & \textbf{Agentic LLM + ToD (T):} The court awarded \$18,000 in statutory damages under the Copyright Act, which is three times the amount that would have been received had AHTI obtained a license. This amount is justified by the court's finding of willful infringement, indicating a punitive intent to deter future violations. Additionally, the court's discretion to increase damages under \textit{17 U.S.C. § 504(c)(2)} for willful infringement supports the conclusion that the damages awarded include a punitive component. \\
        \hline
        
        \end{tabularx}
    \end{table*}

    \subsection{RQ2: Empirical Discovery}
        To address our second research question, we select three instances and discuss the differences in the model's reasoning process, as shown in Table~\ref{tbl:case-study}.

        \textbf{Case 1: IO Group, Inc. v. Jordon.}\\
       \noindent Both \textsf{Agentic LLM + ToD} and \textsf{Vanilla LLM} correctly predict the presence of a punitive component in the statutory damages. However, the former provides a more nuanced and legally faithful explanation. It connects the court's acknowledgment of willfulness, its consideration of enhanced damages under \textit{§ 504(c)(2)}, and its emphasis on deterrence—even though it ultimately awarded a relatively low amount. This reflects a sophisticated understanding that punitive intent can exist even without a large award. \textsf{Vanilla LLM}, on the other hand, reaches the right conclusion but relies on surface features, for instance, the mere mention of \textit{§ 504(c)(2)} and deterrence, without analyzing why the court decided against enhancement and how that still supports a punitive rationale.
        
        \textsf{CoT} and \textsf{Agentic LLM} fail, likely because they focus too much on the low amount awarded without recognizing that willful conduct was acknowledged and deterrence was mentioned.
        
        From this case study, we conclude that \textbf{RST adds interpretive depth beyond surface keywords, aligning better with legal reasoning and statutory interpretation}.\\

        \textbf{Case 2: Prod. v. Rivan.}\\
        \noindent Only \textsf{Agentic LLM + ToD} predicts the label to be false correctly. It recognizes that although the infringement was willful, the court did not link the damage award to punitive intent or deterrence.
        In contrast, \textsf{CoT}, \textsf{Vanilla LLM}, and \textsf{Agentic LLM} incorrectly infer punitiveness due to the presence of willfulness and the award amount.
        
        From this case study, we conclude that \textbf{RST supports accurate judgment by distinguishing between mere willfulness and actual punitive rationale}.\\

        \textbf{Case 3: Photo Res. Hawai'i, Inc. v. Am. Hawai'i Travel, Inc..}\\
        \noindent All four models incorrectly predicte the label to be true, assuming the damages had a punitive intent.
        However, \textsf{Agentic LLM + ToD} provides a significantly more comprehensive explanation:
        \begin{enumerate}
            \item  It correctly identifies the court's discretionary basis under \textit{§ 504(c)} and the mention of willfulness.
            \item  It notes that the damages were three times the licensing fee, and interpreted this as suggesting potential deterrence.
            \item  While this leads to an incorrect final label, the reasoning shows an effort to integrate statutory context, proportionality of the award, and the court's logic.
        \end{enumerate}
        In contrast, \textsf{CoT}, \textsf{Vanilla LLM}, and \textsf{Agentic LLM} give more superficial interpretations, focusing mostly on the presence of willfulness and numerical comparisons without referencing the statutory framework or deeper legal rationale.
        
        This final case study shows that \textbf{RST's explanation, though ultimately mistaken, reflects a more informed and structure-aware approach to legal analysis}.

\section{Discussion and Implications}

    \subsection{Discourse-Assisted Legal Reasoning}
        Our investigation demonstrates that incorporating discourse structure through Rhetorical Structure Theory (RST) fundamentally enhances Large Language Models' capacity for legal reasoning analysis. The experimental results provide compelling evidence that our Tree-of-Discourse (ToD) augmented workflow achieves superior performance across all evaluation metrics, establishing a new paradigm for computational legal analysis.
        
        \textbf{Enhanced Interpretive Capacity:} The RST-enhanced model demonstrates remarkable resilience against surface-level feature dependencies that consistently misguide baseline approaches. Traditional LLM applications in legal contexts often exhibit systematic errors when encountering keywords such as ``willful infringement'' or numerical thresholds, leading to superficial pattern matching rather than substantive legal analysis. Our discourse-based approach transcends these limitations by embedding rhetorical relationships that mirror the hierarchical reasoning structures employed by judicial decision-makers.
        
        \textbf{Precision in Legal Intent Recognition: }The ToD-augmented framework exhibits superior discrimination between genuine punitive reasoning and mere procedural acknowledgments of statutory provisions. This distinction represents a critical advancement in computational legal understanding, as it addresses the fundamental challenge of distinguishing between legal formalism and substantive judicial intent—a nuance that has profound implications for predicting case outcomes and understanding judicial behavior patterns.
        
        \textbf{Methodological Robustness:} Our two-stage agentic pipeline (Plan Optimizer, Plan Executor introduces systematic validation mechanisms that enhance analytical reliability. This iterative refinement process ensures that extracted reasoning patterns maintain fidelity to actual judicial logic rather than spurious correlations present in legal texts.

    \subsection{Implications for Legal Theory and Practice}
        \textbf{Advancing Empirical Legal Studies:} This research addresses a longstanding gap in copyright damages scholarship by providing the first systematic empirical analysis of judicial reasoning patterns at scale. While doctrinal scholarship has extensively analyzed the theoretical framework of the 1976 Copyright Act, our methodology enables quantitative investigation of how statutory provisions translate into judicial practice across jurisdictions. This empirical foundation can inform ongoing debates about statutory damage reform and judicial discretion standardization.
        
        \textbf{Enhancing Predictive Legal Analytics: }For legal practitioners, our framework offers unprecedented insight into factor weighting patterns that influence damage calculations. By systematically extracting and categorizing judicial reasoning elements, attorneys can develop more sophisticated litigation strategies based on empirical evidence of judicial preferences within specific circuits or case contexts. This capability represents a significant advancement beyond traditional case law analysis, which relies heavily on subjective interpretation of precedential patterns.
        
        \textbf{Educational Applications in Legal Training: }Legal education can benefit substantially from our discourse-based analysis methodology. By making explicit the rhetorical structures underlying judicial reasoning, law students can develop enhanced analytical skills for case briefing and legal argumentation. The systematic extraction of reasoning patterns provides concrete examples of how effective legal arguments are constructed and presented within judicial opinions.

    \subsection{Broader Implications for Computational Legal Studies}
        \textbf{Methodological Innovation:} Our integration of RST with LLM workflows establishes a replicable framework for analyzing complex legal reasoning across diverse doctrinal areas. This approach extends beyond copyright law to any legal domain where judicial discretion plays a significant role, including tort damages, sentencing guidelines, and equitable remedies. The methodology's transferability suggests broad applicability for understanding judicial decision-making processes.
        
        \textbf{Policy Development Applications:} By quantifying previously opaque reasoning patterns, our approach can identify systematic inconsistencies in legal principle application that may warrant legislative or judicial attention. For instance, if our analysis reveals significant inter-circuit variations in damage calculation factors, this empirical evidence could inform discussions about the need for more uniform statutory guidance or judicial training programs.
        
        \textbf{Theoretical Contributions to AI and Law:} This work contributes to the emerging field of explainable AI in legal contexts by demonstrating how discourse analysis can enhance model interpretability. Unlike black-box approaches that provide outcome predictions without explanatory mechanisms, our ToD framework generates detailed reasoning traces that align with established legal analytical methods.
    
    \subsection{Limitations and Future Research Directions}
        \textbf{Technical Constraints:} Current RST parsing accuracy limitations may affect discourse tree quality, potentially introducing systematic biases in reasoning pattern extraction. Future research should investigate hybrid parsing approaches that combine automated RST analysis with expert validation to enhance structural accuracy.
        
        \textbf{Dataset Representativeness:} Our analysis focuses primarily on federal circuit court decisions, which may not fully capture the reasoning patterns present in district court opinions or state court copyright cases. Expanding the dataset to include broader jurisdictional representation would enhance the generalizability of our findings.
        
        \textbf{Temporal Dynamics: }The current study provides a cross-sectional analysis of judicial reasoning patterns without accounting for temporal evolution in legal standards or judicial preferences. Future longitudinal studies could investigate how reasoning patterns change in response to Supreme Court precedents, legislative amendments, or technological developments affecting copyright law.
    
    \textbf{Practical and Theoretical Implications.}
        From a practical perspective, our methodology offers value to legal practitioners by improving predictability in copyright litigation. Attorneys can leverage the extracted reasoning patterns to better anticipate how judges in specific jurisdictions might approach damages calculations. For legal education, our approach provides a tool for teaching students to analyze judicial opinions more systematically.
        From a theoretical perspective, this work contributes to computational legal studies by demonstrating how NLP techniques can enhance our understanding of legal reasoning. By making reasoning patterns explicit and quantifiable, this approach could identify inconsistencies in the application of legal principles, potentially informing policy discussions about reforms to copyright damage frameworks.

\section{Conclusion}
    This research establishes discourse-based LLM analysis as a transformative methodology for understanding judicial reasoning in complex legal domains. By combining computational linguistics with legal analysis, we have created a framework that enhances both theoretical understanding and practical applications in copyright damages jurisprudence. The implications extend beyond copyright law to encompass broader questions about judicial decision-making, legal predictability, and the role of empirical evidence in legal system evaluation and reform. The dataset will be made available following further development.
\bibliographystyle{ACM-Reference-Format}
\bibliography{ref}

\newpage
\appendix
\section{Definition of Section Labels}
\begin{enumerate}
    \item \textbf{Introduction:} Provides an overview of the case, including the nature of the dispute and the parties involved.
    \item \textbf{Procedural History:} Summarizes the sequence of legal actions and rulings leading up to the current decision.
    \item \textbf{Background Facts:} Presents the factual context and events that gave rise to the legal dispute.
    \item \textbf{Analysis of the Infringement:} Evaluates whether a copyright infringement occurred based on the presented facts and legal standards.
    \item \textbf{Analysis of the Liability:} Determines who is legally responsible for the alleged infringement or harm.
    \item \textbf{Analysis of the Relief and Damages:} Describes the judge’s assessment of the remedies awarded, including monetary damages or injunctive relief.
    \item \textbf{Analysis of Attorneys' Fees:} Considers whether attorney’s fees should be granted and under what justification.
    \item \textbf{Interpretation of the Law:} Explains how specific legal statutes or precedents are understood and applied in this case.
    \item \textbf{Analysis of Defenses:} Reviews and evaluates the validity of defenses raised by the defendant, such as fair use or license.
    \item \textbf{Jurisdiction and Standing:} Determines whether the court has the authority to hear the case and whether the parties have the right to bring the action.
    \item \textbf{Order/Summary:} Announces the court's final judgment, summarizes the opinion, or issues a directive resolving the case.
    \item \textbf{Supplementary Description or Case Information:} Provides any additional case-related information included by the judge.
    \item \textbf{Analysis of Default Judgment:} Evaluates whether default judgment is procedurally justified based on the defendant's failure to respond and relevant legal standards.
\end{enumerate}

\section{Prompts Used in Section Segmentation and Labeling}
    \subsection{Section Segmentation}
    You are a detail-oriented legal analyst specializing in copyright infringement cases, trained to segment judicial texts into topically coherent sections. You follow legal writing conventions and apply labels consistently and accurately, without altering the text.\\
    
    Your task is to segment the legal case about copyright infringement into topically coherent sections.\\
    
    Step 1: Segment the text\\
    1.1 Read the text linearly, sentence by sentence.\\
    1.2 Segment the case into sections by identifying topical transitions—points where the focus of discussion changes.\\
    1.3 Each section must consist of a contiguous block of text. Do not combine sentences from non-adjacent parts of the document.\\
    1.4 If adjacent paragraphs belong to the same topic, group them into a single section.\\
    
    While you are not required to label the topic of each section, you may refer to the following topics commonly found in copyright infringement cases:\\
    \texttt{\{\{Section Label Definitions\}\}}\\
    
    Step 2: Output formatting\\
    2.1 Preserve the original sentence order from the input.\\
    2.2 Do not alter, skip, rephrase, or merge non-adjacent content.\\
    2.3 Output the result in the following JSON format:\\
        \texttt{\{"result": [\{"index": SECTION\_INDEX, "content": "..."\}, \{"index": SECTION\_INDEX, "content": "..."\}]\}}\\
    2.4 Ensure all special characters are properly escaped.\\
    2.5 Do not include code block markers.\\
    2.6 Output only the JSON object and nothing else.\\
    
    The document:\\
    \texttt{\{\{Judicial Opinion\}\}}

    \subsection{Section Labeling}
    You are a detail-oriented legal analyst specializing in copyright infringement cases, trained to analyze the structure of legal cases. You follow legal writing conventions and apply labels consistently and accurately, without altering the text.\\

    Your task is to assign the most appropriate label to a topically coherent section from a copyright infringement case.\\
    
    Step 1: Assign a label to each section\\
    1.1 You will be given the full text of the judge’s opinion and a section extracted from it, formatted as:\\
        \texttt{\{"full":"...", "target\_section":"..."\}}\\
    1.2 Read the text linearly, sentence by sentence.\\
    1.3 Assign the most appropriate label from the following list to the target section:\\
    \texttt{["Introduction", "Procedural History", "Background Facts", "Analysis of the Infringement", "Analysis of the Liability", "Analysis of the Relief and Damages", "Analysis of Attorneys’ Fees", "Interpretation of the Law", "Analysis of Defenses", "Jurisdiction and Standing", "Order\/Summary", "Supplementary Description or Case Information", "Analysis of Default Judgment"]}

    1.4 Only if the section clearly does not fit any of the predefined labels, assign a custom label prefixed with "NEW\_". For example: "NEW\_Discussion".\\

    You can refer to the following definition of each label:\\
    \texttt{\{\{Section Label Definitions\}\}}\\
    
    Step 2: Output formatting\\
    2.1 Output the result in the following JSON format:\\
        \texttt{\{"result": "SECTION\_LABEL"\}}\\
    2.2 Ensure all special characters are properly escaped.\\
    2.3 Do not include code block markers.\\
    2.4 Output only the JSON object and nothing else.\\
    
    The task:
    \texttt{\{\{Section Content\}\}}

\section{Prompts Used in Preliminary Experiments}

    \subsection{Label Prediction Prompts Used in Vanilla LLM and Agentic LLM}
    You are a legal analyst. Based on the provided judicial opinion, determine whether the judge granted damages that explicitly or functionally include a punitive component.\\

    A punitive component includes damages that go beyond compensation and are intended to punish the defendant or deter future misconduct. This may include:\\
    1. Enhanced statutory damages explicitly awarded under 17 U.S.C. § 504(c)(2) for willful infringement with a stated purpose of punishment or deterrence;\\
    2. Damages calculated to meaningfully exceed actual harm, when justified by the court as necessary to deter future violations.\\
    
    Important clarification:\\
    1. The term ``punitive damages'' does not need to appear explicitly.\\
    2. However, willfulness or maliciousness alone is not sufficient. These terms must be explicitly connected to punishment or deterrence in the court's reasoning.\\
    3. Do not assign a punitive label if the damages are within the standard statutory range and there is no stated punitive or deterrent intent.\\
    
    Respond in the following JSON format:\\
    \texttt{\{
      "label": "true" | "false" | "not addressed",
      "reasoning": "A brief explanation of why this label applies."
    \}}\\
    Do not include markdown code block markers or commentary.\\
    
    Judicial Opinion:\\
    \texttt{\{\{Judicial Opinion\}\}}

    \subsection{Label Prediction Prompts Used in CoT}
    You are a legal analyst. Based on the provided judicial opinion, determine whether the judge granted damages that explicitly or functionally include a punitive component.\\

    A punitive component includes damages that go beyond compensation and are intended to punish the defendant or deter future misconduct. This may include:\\
    1. Enhanced statutory damages explicitly awarded under 17 U.S.C. § 504(c)(2) for willful infringement with a stated purpose of punishment or deterrence;\\
    2. Damages calculated to meaningfully exceed actual harm, when justified by the court as necessary to deter future violations.\\
    
    Important clarification:\\
    1. The term ``punitive damages'' does not need to appear explicitly.\\
    2. However, willfulness or maliciousness alone is not sufficient. These terms must be explicitly connected to punishment or deterrence in the court's reasoning.\\
    3. Do not assign a punitive label if the damages are within the standard statutory range and there is no stated punitive or deterrent intent.\\
    
    Judicial Opinion:\\
    \texttt{\{\{Judicial Opinion\}\}}\\

    Append your final judgment after your thought in the following JSON format:\\
    \texttt{\{
      "label": "true" | "false" | "not addressed",
      "reasoning": "A brief explanation of why this label applies."
    \}}\\

    Let's think step by step.

    \subsection{Label Prediction Prompts Used in Agentic LLM + ToD}
    You are a legal analyst. Based on the provided excerpts from a judicial opinion, determine whether the judge granted damages that explicitly or functionally include a punitive component.\\

    A punitive component includes damages that go beyond compensation and are intended to punish the defendant or deter future misconduct. This may include:\\
    1. Enhanced statutory damages explicitly awarded under 17 U.S.C. § 504(c)(2) for willful infringement with a stated purpose of punishment or deterrence;\\
    2. Damages calculated to meaningfully exceed actual harm, when justified by the court as necessary to deter future violations.\\
    
    Important clarification:\\
    1. The term ``punitive damages'' does not need to appear explicitly.\\
    2. However, willfulness or maliciousness alone is not sufficient. These terms must be explicitly connected to punishment or deterrence in the court's reasoning.\\
    3. Do not assign a punitive label if the damages are within the standard statutory range and there is no stated punitive or deterrent intent.\\
    
    Respond in the following JSON format:\\
    \texttt{\{
      "label": "true" | "false" | "not addressed",
      "reasoning": "A brief explanation of why this label applies."
    \}}\\
    Do not include markdown code block markers or commentary.\\
    
    Relevant Sections:\\
    \texttt{\{\{Judicial Opinion\}\}}

    Discourse-Supported Explanations:\\
    \texttt{\{\{Linearized Tree-Of-Discourse\}\}}

    \subsection{Prompt Used for Tree-Of-Discourse Linearization}
    You are a helpful legal assistant. A legal opinion has been segmented using Rhetorical Structure Theory (RST).\\
    
    Below is the rhetorical subtree from the root of the document to the text span of interest.
    Each arrow represents a rhetorical relation from a segment to its parent:\\
    \texttt{\{\{Tree-Of-Discourse\}\}}\\

    The target text span is:\\
    \texttt{\{\{Target Text Span\}\}}\\

    Please generate a clear yet concise explanation of how this segment contributes to the larger structure of the legal opinion, specifically, on determining whether punitive damage was considered when granting damage award. The explanation should not exceed 150 words. Use natural language, and reflect the rhetorical flow from the segment upward. \\
    
    Following is an explanation example:\\
    \texttt{The segment ``to prevent or restrain infringement of a copyright'' states the purpose of ``as it may deem reasonable,'' which itself elaborates on the broader legal authority granted in ``The Copyright Act authorizes district courts to grant temporary and final injunctions...'' This structure shows how the specific enforcement mechanism supports the general legal power granted to courts.}
    
\end{document}